# Hardware support for arbitrarily complex loop structures in embedded applications

Nikolaos Kavvadias and Spiridon Nikolaidis
*Section of Electronics and Computers, Department of Physics*
*Aristotle University of Thessaloniki, 54124 Thessaloniki, Greece*
{*nkavv@skiathos.physics.auth.gr*}

**Abstract**

*In this paper, the program control unit of an embedded RISC processor is enhanced with a novel zero-overhead loop controller (ZOLC) supporting arbitrary loop structures with multiple-entry/exit nodes. The ZOLC has been incorporated to an open RISC processor core to evaluate the performance of the proposed unit for alternative configurations of the selected processor. It is proven that speed improvements of 8.4% to 48.2% are feasible for the used benchmarks.*

## 1 Introduction

Last years, the embedded processor market is dominated by new 32-bit RISC architectures (ARM, MIPS32), and embedded DSPs (Motorola 56300, ST120, TMS320C54x) featuring architectural and power consumption characteristics suitable to portable platforms. Some of these new features provide better means for the execution of loops, by surpassing the significant overhead of the *loop overhead instruction pattern* which consists of the required instructions to initiate a new iteration of the loop.

Generally, looping cycle overheads are confronted by using branch-decrement instructions, zero-overhead loops or customized units for more complex loop nests [1], [2]. For the XiRisc processor [1], branch-decrement instructions can be configured prior synthesis. For the DSP56300 supporting up to 7 nesting levels there is a 5-cycle overhead applied even to the innermost loops. Also, a single-cycle multiple-index update unit for perfect loop nests has been described in [2]. Its main advantage is that successive last iterations of nested loops are performed in a single cycle. In contrast to our approach, only perfect loop nests are supported and the area requirements grow proportionally to the considered number of loops.

In our approach, a ZOLC method is introduced that eliminates the loop overheads and can be applied to an arbitrary combination of loops. The initialization of ZOLC presents only a very small cycle overhead since it occurs outside of loop nests. The unit has been incorporated to the XiRisc 32-bit processor [1], which is distributed as a VHDL soft-core.

## 2 Incorporating ZOLC to a programmable processor

A block diagram indicating how the proposed ZOLC architecture is incorporated in the control path of a typical RISC processor is shown in Fig. 1. The purpose of ZOLC is to provide a proper candidate program counter (PC) target address to the PC decoding unit for each substituted looping operation. The instruction decoder, the PC decoding unit and the general-purpose register file communicate with the ZOLC hardware. ZOLC is composed from the task selection unit, which determines the appropriate next PC value when execution resides in a loop structure, the loop parameter tables where the loop bound values are kept and the index calculation unit.

Two modes of operation are distinguished for the ZOLC. In "initialization" mode, the ZOLC storage resources are initialized with the known loop bound values and the loop structure encoding by a special instruction sequence. In "active" mode, the ZOLC: a) determines the following task, b) issues a new target PC value and a set of candidate exit values for the case of multiple-exit loops to PC decode, c) loop indices are updated and written back to the integer register file. The task sequencing information (tasks are defined as control-flow graph regions among loop boundaries) is stored in a LUT within the task selection unit. On completion of a task, a task end signal is issued from PC decode, and an entry is selected from the LUT to address the succeeding task and the loop parameter blocks, based on which task has completed and the current loop status.

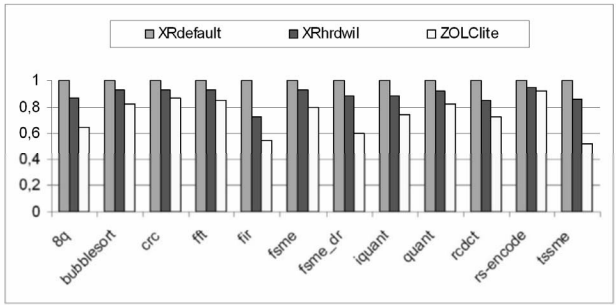

**Figure 2. Cycle performance results for the examined applications**

## 3 Performance Evaluation of ZOLC

We have implemented three different configurations of a ZOLC engine. *ZOLCfull* refers to a ZOLC supporting 32 task switching entries, and 8-loop structure with up to 4 entries/exits/entries per loop. *ZOLClite* lacks support for multiple-entry/exit and *uZOLC*, is usable for single loops. Along with the three variations of ZOLC, two instances of the XiRisc processor family are invoked, the unmodified core noted as *XRdefault*, and *XRhrdwil* employing branch-decrement instructions.

It was found for *uZOLC*, *ZOLClite* and *ZOLCfull*, that the requirements in storage resources are 30, 258 and 642 storage bytes and in combinational area 298, 4056, and 4428 equivalent gates, respectively. The processor cycle time is not affected due to ZOLC and corresponds to about 170MHz on a 0.13μm ASIC process. The relative cycle measurements of Fig. 2 for comparing *ZOLClite* against two XiRisc configurations have been acquired for a set of 12 benchmark applications, collected from the XiRisc validation suite [1], and software implementations of motion estimation kernels. The use of branch-decrement instructions provides a cycle reduction of up to 27.5% and about 11.1% in average, while incorporating the ZOLC unit is responsible for improvements of up to 48.2% and about 26.2% in average.

## 4 Conclusion

In this paper, a zero-overhead loop controller is introduced and incorporated to the XiRisc processor. The presented architecture is able to execute structured algorithms for an arbitrary combination of loops, with no cycle overheads incurred for task switching. For a representative benchmark suite, execution time improvements up to 48% are reported against alternative XiRisc configurations.

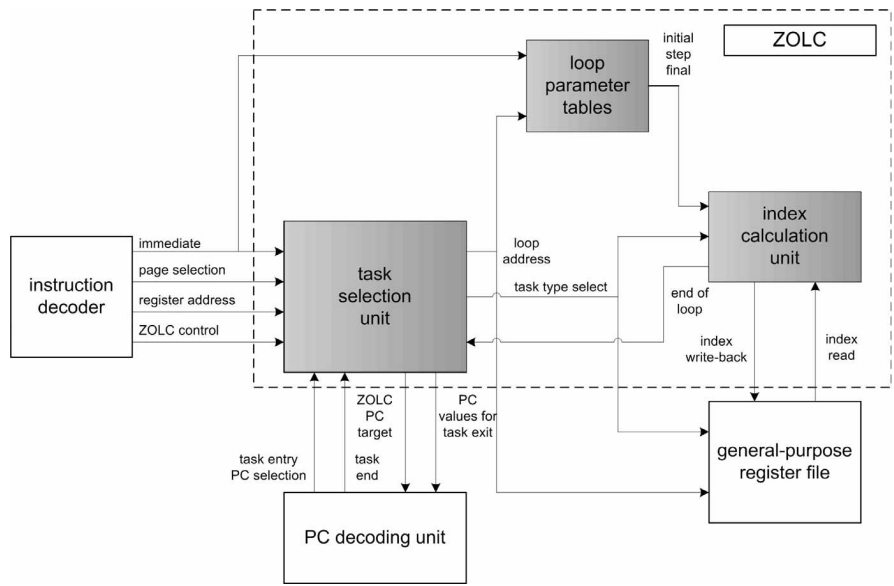

**Figure 1. Incorporating the ZOLC architecture to programmable RISC processors**